\documentclass[twocolumn,showpacs,prl]{revtex4}

\usepackage{graphicx}
\usepackage{dcolumn}
\usepackage{bm}

\begin{document}


\title{Anomalous Suppression of Superfluidity in $^4$He Confined in a Nano-porous Glass:\\Possible Quantum Phase Transition}

\author{K. Yamamoto}
\author{H. Nakashima}
\author{Y. Shibayama}
\author{K. Shirahama}%
 \email{keiya@phys.keio.ac.jp}
\affiliation{%
Department of Physics, Keio University, Yokohama 223-8522, Japan
}%

\date{\today}

\begin{abstract}
We explore superfluidity for $^4$He confined in a porous glass which has nanopores of 2.5 nm in diameter, at pressures up to 5 MPa. With increasing pressure, the superfluidity is drastically suppressed, and the superfluid transition temperature approaches 0 K at $P_c = 3.5$ MPa. The features strongly suggest that the extreme confinement of $^4$He into the nanopores induces a quantum phase transition from superfluid to nonsuperfluid at 0 K, and at $P_c$. 
\end{abstract}

\maketitle
$^4$He adsorbed or confined in porous media provides excellent examples of interacting Bose system. The system dimensionality and interatomic interaction can be easily controlled by changing pore size, pore structure and $^4$He density\cite{Reppy,Shirahama,Wada}. In addition, disorder in the porous structures results in essential changes in the properties such as superfluid critical phenomena\cite{Chan}.

In the case of $^4$He confined in narrow pores of 10 nm in size, the pressure - temperature ($P-T$) phase diagram is altered: the freezing is inhibited and the superfluidity is slightly suppressed. For $^4$He in a porous Vycor glass, which has randomly and three-dimensionally (3D) connected pores of 6 nm in diameter, the freezing pressure at about 1.5 K increases to 4 MPa and the superfluid $\lambda$ line shifts about 0.2 K from that of bulk $^4$He (see Fig. \ref{Fig3}(a))\cite{Adams,Beamish,Lie-Zhao,Bittner}. These changes are attributed to the inhibition of crystal nucleation in narrow pores and the suppression of the superfluid order parameter near the pore walls. 

As the suppression of superfluidity is enhanced with decreasing pore size\cite{Lie-Zhao}, an interesting question arises: if the pore size is extremely decreased, how the superfluidity is suppressed? In this Letter, we demonstrate that the confinement of $^4$He into a nano-porous medium leads to an unexpectedly strong suppression of superfluidity, resulting in a qualitative change in the phase diagram.

We study the pressure effects on superfluid $^4$He confined in a porous Gelsil glass\cite{Gelsil}. The structure of Gelsil is characterized by a 3D random network of nanopores, similarly to Vycor. The present Gelsil sample, however, has much smaller pores, 2.5 nm in nominal diameter. The superfluidity of the adsorbed $^4$He films in a similar Gelsil sample was studied by Miyamoto and Takano\cite{Miyamoto}, and $T_c$ for $^4$He filled in the glass was found to be 0.9 K. Based on this suppression of $T_c$ at saturated vapor pressure, a further suppression is expected at higher pressures. 

The Gelsil sample we use is a disk of 5.5 mm in diameter, 2.5 mm thick, and 55.25 mg in weight. We heated the glass up to 150 $^{\circ}$C in vacuum to remove adsorbed molecules, especially water in the pores. The surface area measured by N$_2$ adsorption is 26.9 m$^2$.
We measure the superfluid response employing a torsional oscillator. It consists of a brass sample cell containing the glass disk, and a Be-Cu hollow torsion rod, which acts as a $^4$He filling line. In order to exert an uniform pressure to the liquid $^4$He in the glass sample, an open space of 0.5 mm thick is remained between the glass and the hollow region of the rod.
The cell oscillates at a frequency $f \sim 1955.4$ Hz, with a high quality factor, e.g. $Q \sim 4 \times 10^6$ at 10 mK. It is cooled to 9 mK with a dilution refrigerator. 

We measure the temperature dependence of the frequency shift $\Delta{f}(T)$, which is proportional to the superfluid density $\rho{_s}(T)$, and the change in the oscillation amplitude (as a voltage $V(T)$), in a very wide range of $^4$He density. The feed of $^4$He into the cell and the density (pressure) control are made by a room-temperature gas handling system. When the amount of the fed $^4$He is not enough to fill the nanopores, $^4$He forms a thin film adsorbed on the pore walls. We refer to such a situation as the film state. If $^4$He is continuously fed, on the other hand, the liquid $^4$He fills not only the nanopores but also the open space in the cell, the hole in the torsion rod, and the Cu-Ni inlet tube. In this situation we can control the liquid pressure in the nanopores with the gas handling system. This situation is referred to as the pressurized state. The pressure is monitored at room temperature.

\begin{figure}[tbp]
  \begin{center}
     \includegraphics[keepaspectratio=true,height=130mm]{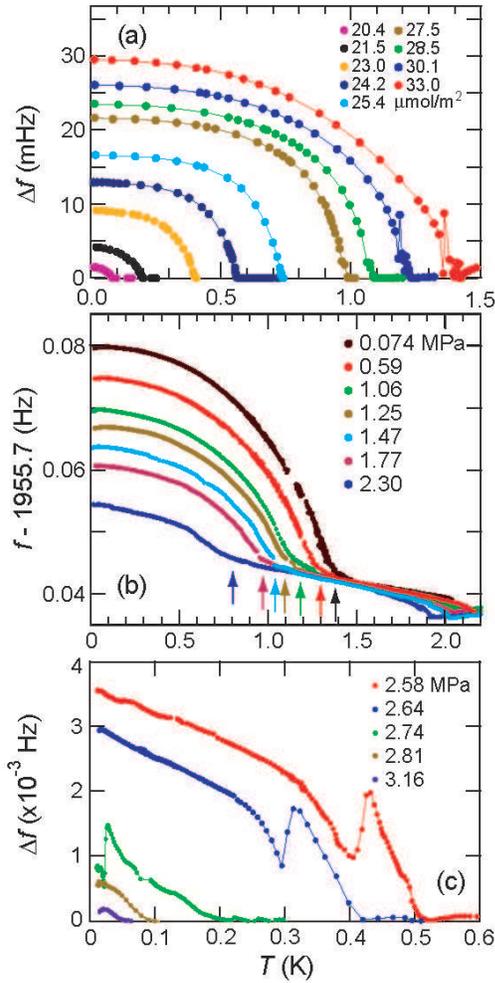}
       \end{center}
  \caption{(a) $\Delta{f}(T)$ in the adsorbed film states for various coverages, from $n = 20.4$ to $33.0 \mu$ mol/m$^2$. (b) $f(T)$ in the pressurized states. To clarify the superfluid transitions, the frequencies are shifted so as to collapse onto a single curve between 1 and 1.5 K. The frequency value is valid for the data of 0.074 MPa, otherwise shifted. Arrows indicate $T_c$ in the nanopores. (c)  $\Delta{f}(T)$ above the bulk freezing pressure 2.53 MPa for various $P_{cell}$'s estimated by the procedure described in the text. The "$n$ - shape" anomalies seen in the data in (a) and (c) are caused by resonant couplings to the superfluid third or fourth sounds.}
    \label{Fig1}
\end{figure}

We observe distinct superfluid transitions both in the film states and in the pressurized states, as shown in Fig. \ref{Fig1}. In the film states we show $\Delta{f}(T)$ in Fig. \ref{Fig1}(a), and the dependence of the extracted $T_c$ on the coverage $n$ in Fig. \ref{Fig3}(b). Up to about 20 $\mu$mol/m$^2$, denoted as a critical coverage $n_c$, no superfluidity is observed due to the strong van der Waals attraction from the glass wall. As $n$ increases, the superfluid film grows on the nonsuperfluid layers, and $T_c$ increases nearly linearly. These features are common in $^4$He films adsorbed on various porous substrates, and are possibly a manifestation of interesting natures in both 2D and 3D superfluidity\cite{Shirahama,Reppy}.  

At $n=33 $ $\mu$mol/m$^2$, $T_c$ reaches maximum, 1.43 K. Above this coverage, we observe in $f(T)$ a contribution from bulk $^4$He accumulated in the open space of the cell, indicating that the pores are filled with liquid $^4$He.
Then $T_c$ slightly decreases and tends to 1.35 K. The overall behaviors are qualitatively similar to those observed by Miyamoto and Takano\cite{Miyamoto}. However,
our maximum $T_c$ is about 1.5 times higher than that observed by them. Moreover, there are substantial differences in the $\Delta{f(T)}$ curves. We guess that the average pore size in our Gelsil sample is slightly larger than that of theirs.

As more $^4$He is added, the pressurized state is achieved. The $^4$He density in the nanopores is thus controlled by pressurizing the $^4$He gas at room temperature. The presence of bulk liquid in the cell, however, causes large ambiguity in the estimation of $\Delta{f}(T)$. So we show in Fig. \ref{Fig1}(b) $f(T)$ for various pressures, by shifting the frequencies so that the data above $T_c$ collapse onto a single curve. The abrupt frequency changes appearing at about 2 K are due to the superfluid transition of bulk $^4$He in the cell. We show $T_c$ for various pressures in Fig. \ref{Fig3}(a). It decreases monotonically with increasing pressure, and reaches 0.71 K at 2.47 MPa, just below the bulk freezing pressure. It is remarkable that the pressure dependence of $T_c$, $\left|dT_{c}/dP\right|$, increases progressively with increasing $P$. 

\begin{figure}[tbp]
  \begin{center}
    \includegraphics[keepaspectratio=true,height=85mm]{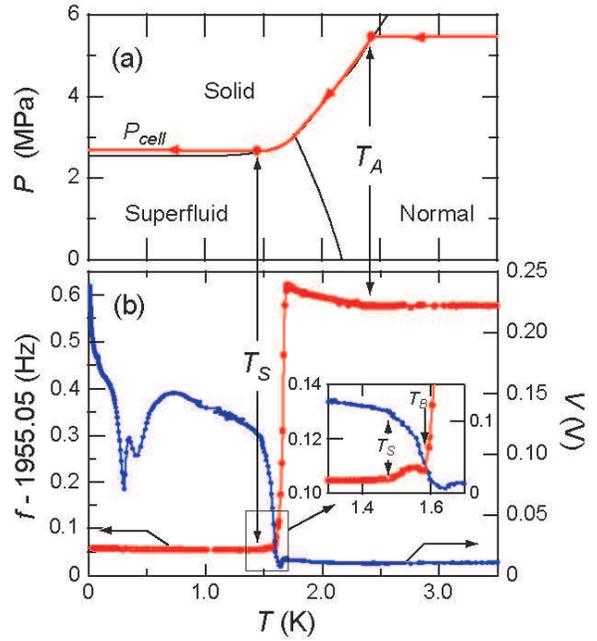}
  \end{center}
  \caption{Estimation of $P_{cell}$ over 2.53 MPa. (a) The bulk phase diagram. The red line shows $P_{cell}$, tracing the "cooling path" of the system including the cell and the filling tube. After reaching the bulk freezing curve at $T_A$, $P_{cell}$ rapidly decreases along with the freezing curve (the liquid-solid coexistence state). At $T_S$, the solidification of $^4$He in the cell is completed, and $P_{cell}$ below $T_S$ is set at the freezing pressure at $T_S$. (b) Changes in $f(T)$ and the oscillator amplitude $V(T)$ during the cooling of the cell. Two minima in the amplitude below 0.5K are caused by the fourth sound resonances in $^4$He inside the pores. Inset: enlarged plot near $T_B$ and $T_S$. For explanation, see text.}
    \label{Fig2}
\end{figure}

Above the bulk freezing pressure 2.53 MPa, $^4$He in the nanopores still keeps liquid state and exhibits superfluidity, while bulk $^4$He solidifies. The pressure in the sample cell cannot be controlled from room temperature, because the bulk $^4$He in the cell and the filling tube solidifies and blocks pressure change. We estimate the pressure in the cell $P_{cell}$ above 2.53 MPa, from the behaviour of $f(T)$ and $V(T)$ during cooling. For explanation we show the bulk phase diagram in Fig. \ref{Fig2}(a), and a typical example of $f(T)$ and $V(T)$ in Fig. \ref{Fig2}(b). We start cooling the cell from 3.5 K with keeping the pressure at room temperature, $P_{RT} = P_{cell} = 5.5$ MPa.  The frequency starts to increase at $T_{A} = 2.43$ K, where the refrigerator reaches the freezing curve. This indicates that liquid $^4$He starts to solidify somewhere in the filling tube above the still of the dilution refrigerator (our refrigerator has no 1 K pot). The increase in $f(T)$ is caused by the decrease in the $^4$He density in the cell. Below $T_A$, the filling tube is blocked by solid $^4$He, and $P_{cell}$ is no longer equal to $P_{RT}$. $P_{cell}$ decreases along with the freezing curve, keeping the liquid-solid coexistence. During cooling, the solid plug grows toward lower temperature.

At 1.65 K, $f(T)$ abruptly decreases
, and $V(T)$ starts to increase. We attribute these changes to the solidification of $^4$He in the open volume of the cell and in the torsion rod. The solid $^4$He couples perfectly to the torsion oscillation, whereas the normal liquid does not. This difference in the oscillator response results in the decrease in $f(T)$ and the increase in $V(T)$. After the large decrease (0.57 Hz), $f(T)$ slightly increases at a temperature denoted as $T_B = 1.590$ K, and shows a hump, and eventually stops to decrease at $T_S = 1.477$ K, as shown in the inset of Fig. \ref{Fig2}(b). The origin of the hump is not understood. At the same time, $V(T)$ abruptly increases between $T_B$ and $T_S$. Below $T_S$, $f(T)$ and $V(T)$ trace the same temperature dependence as those of the empty cell. We therefore conclude that the solidification of bulk $^4$He is completed at $T_S$. The liquid $^4$He inside the nanopores coexists with the bulk solid $^4$He in the cell below $T_S$. Since the solid pressure stays nearly constant at all temperatures below $T_S$, the liquid pressure inside the pores also stays at the melting pressure at $T_S$, i.e.  $P_{cell}(T < T_S) = P_{cell}(T_S)$. 
We note that the behaviors in $f(T)$ and $V(T)$ are reproducible, and the overall behaviors are independent of the initial pressure $P_{RT}$. 

The superfluid transition is clearly observed at $P_{cell} > 2.53$ MPa. In Fig. \ref{Fig1}(c), we show $\Delta{f}(T)$ for various $P_{cell}$'s determined by the procedure described above. As $P_{cell}$ increases, both $T_c$ and $\Delta{f(0)}$ monotonically decrease. At $P_{cell} = 3.45$ MPa, we find that $T_c = 38.1$ mK. As it is difficult to tune finely $P_{cell}$, it is the lowest $T_c$ that we have observed so far. At pressures ranging from 3.9 MPa to 5.0 MPa we have not detected the superfluid signals.

All the $T_c$ data obtained are plotted in the phase diagram shown in Fig. \ref{Fig3} together with that for the film states. We find that $T_c$ approaches 0 K at pressure $P_c \sim 3.5$ MPa. This feature is radically different from that in bulk $^4$He or in $^4$He confined in Vycor, in which the "$T_c$ line" terminates at the freezing curve as shown in the same figure. Note that the superfluid phase exists in the limited range of density, from $n_c$, to $P_c$, which can be alternatively expressed as a second critical density. 

It is difficult to obtain $\Delta{f}(0)$ with a high precision, as $f(T)$ is unknown below 9 mK. We show $\Delta{f}(10 {\rm mK})$ in place of $\Delta{f}(0)$, in Fig. \ref{Fig4}. It decreases continuously to zero as $P_c$ is approached.
Unfortunately, large scatters in the data prevent to determine their functional dependences (e.g. powerlaw) on pressure. The scatters, however, are caused only by the pressure estimation. It is consequently definite that both $T_c$ and $\rho_{s}(0)$ approach zero.
\begin{figure}[tbp]
  \begin{center}
    \includegraphics[keepaspectratio=true,height=100mm]{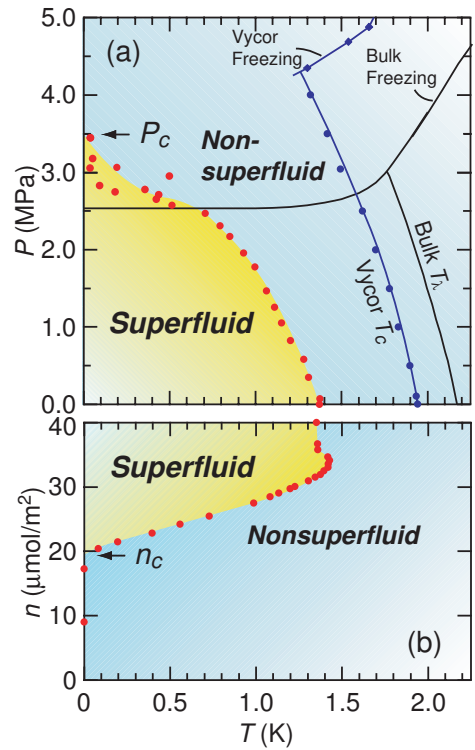}
  \end{center}
  \caption{(a) The $P-T$ phase diagram. The yellow area shows the superfluid phase in Gelsil. The phase boundaries of bulk $^4$He and $^4$He in a porous Vycor glass\cite{Lie-Zhao} are also shown. (b) The phase diagram for the film states.}
      \label{Fig3}
\end{figure}

The continuous suppressions in $T_c$ and $\rho_{s}(0)$ to zero are quite unprecedented for superfluid $^4$He in restricted geometries. In any other confined geometries previously studied\cite{Adams,Beamish,Lie-Zhao,Bittner}, the changes of the phase diagram were only quantitative; i.e. the $T_c$ line shifts in parallel with the bulk $\lambda$ line. The superfluid suppression in Gelsil cannot be attributed to the ordinary superfluid size effect. Moreover, contrary to the bulk superfluid-solid transition, the present superfluid-nonsuperfluid transition at 0 K driven by pressure is not first order, but continuous.

The decrease in $\rho_s(0)$ with increasing $P$ was also observed in $^4$He in Vycor, and it was attributed to blockade of pores caused by local solidification of $^4$He\cite{Lie-Zhao}. If the solidification took place locally in some pores of the glass sample, the superflow would be blocked there, resulting in the decrease in $\Delta{f(T)}$. Contrary to our observation, however, $T_c$ would not decrease, because such solid plugs do not affect the genuine superfluid density. Therefore the suppressions in both $T_c$ and $\rho_s(0)$ in Gelsil are not due to the classical blockade effect, but due to the essential change in the nature of superfluidity. This conclusion is reinforced by the observations of coupled fourth-sound resonances, which are seen as "$n$-shape" anomalies in $\Delta{f}(T)$ of Fig. \ref{Fig1}(c), and as two minima in $V(T)$ shown in Fig. \ref{Fig2}(b). The existence of sound resonances ensures that the superfluidity takes place in a macroscopic scale.
 
All the results mentioned above strongly suggest that $^4$He confined in the nanopores of Gelsil undergoes a continuous quantum phase transition (QPT) at $P_c$ and at 0 K, in the sense that the continuous (non first order) superfluid-nonsuperfluid transition is driven by changing pressure as an externally controllable parameter\cite{Sondhi}. For further confirmation of the QPT, the nature of the nonsuperfluid state near $P_c$ needs to be elucidated. We have found no indications of solidification up to $P_{cell}=5$ MPa. The torsional oscillator may not have sensitivity for the solidification in the nanopores. Bittner and Adams measured the $^4$He freezing curve in other porous glass of 2.4 nm pore size down to 1.4 K. It was nearly the same as that in Vycor\cite{Bittner}. Even if we assume that our $^4$He-Gelsil system has the similar freezing curve, it is difficult to predict the behavior below 1 K. We currently prepare a simultaneous measurement of pressure, ultrasound and torsional oscillator to reveal the complete phase diagram. Heat capacity measurement will be also useful for understanding the nature of both phases.

\begin{figure}[tbp]
  \begin{center}
    \includegraphics[keepaspectratio=true,height=30mm]{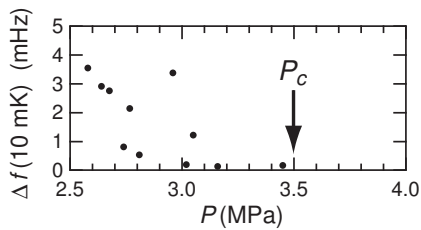}
  \end{center}
  \caption{$\Delta{f}$ at 10 mK, as a function of $P_{cell}$ above 2.53 MPa.}
  \label{Fig4}
\end{figure}

It should be noted that evidences for QPT have been found around the critical coverage $n_c$ of $^4$He films adsorbed on Vycor\cite{Crowell}. A theory\cite{Fisher} has predicted that disorder in such porous media results in a QPT between the superfluid and the gapless Bose glass phase at $n_c$, although the heat capacity data was not consistent with the Bose glass nature\cite{Crowell}. The $^4$He-Gelsil system may be characterized by two QPTs, at $n_c$ and at $P_c$.

Smallness and disorder of the pore structure may be responsible for the reduction of $T_c$ and  $\rho_s{(0)}$. Since about 1.5 $^4$He atomic layers on the pore walls ($n_c \sim20\mu$mol/m$^2$) are nonsuperfluid, the "real" pore diameter for superfluid is about 1.5 nm, assuming that the effective size of $^4$He atom is from 0.3 to 0.4 nm. The number of the superfluid atoms in the cross section of the 1.5 nm diameter pores is estimated to be about 20. This is about an order of magnitude smaller than that estimated for $^4$He in Vycor ($\sim 200$ atoms for 6 nm pore size). Due to this extreme smallness, the positional exchanges between $^4$He atoms, in particular the long cyclic permutations, which are necessary for possessing superfluidity\cite{Pollock}, may be greatly suppressed. This restriction in atomic exchange may result in the reduction of $T_c$ down to zero. Moreover, the permutations of atoms can be further disturbed by the presence of disorder in the pore structure. In short, the $^4$He atoms can localize in the pores by correlation and disorder. 

Finally, we comment on the dependence of $\Delta{f(T)}$ on $^4$He density. From Fig. \ref{Fig1} we see that the transition in the films is sharper than that in the pressurized states, and in the pressurized states it becomes smeared as $P$ increases. This suggests that the effects of the inevitable pore-size distribution, roughness in the pore walls, and structural disorder on the superfluid density, depend on $^4$He density and the state (film or pressurized). Moreover, assuming a powerlaw for $\Delta{f(T)}$ at low temperatures, i.e. $\Delta{f(T)} = \Delta{f(0)} - const\cdot T^{\alpha}$, the exponent $\alpha$ is found to change from 2.5 at 0.074 MPa, to 1 at 2.64 MPa. This is interpreted as the change in dimensionality of phonons, from one (or more) to zero dimension\cite{Bishop}. Excitation studies such as neutron scattering\cite{Plantevin} under pressure will provide more detailed information. 

In conclusion, we observe the strong superfluid suppression induced by the extreme confinement of $^4$He into the nanopores. The results cannot be explained in terms of the conventional concepts of superfluid coherence length and size effect, but give a strong evidence for a novel type of quantum phase transition. $^4$He-nanopore systems will be important for pursuit of general problems in strongly correlated bosons in a disordered potential.  

This work is supported by the Sumitomo foundation.


\end{document}